\documentclass[superscriptaddress,twoside,twocolumn,prl,floatfix,letterpaper]{revtex4}

\usepackage{graphicx} 
\usepackage{amsmath}
\usepackage{amssymb}
\usepackage{fancyhdr}
\usepackage{journals}
\usepackage{rotating}
\newcommand{\href}[2]{#2}                   

\def \Vista {{\sc Vista}}
\def \Sleuth {{\sc Sleuth}}

\def \DZero {{D0}}

\def \pmiss {{\ensuremath{{\,/\!\!\!p}}}}

\def \SumPt {{\ensuremath{\sum{p_T}}}}

\def \scriptP {\ensuremath{{\cal P}}}
\def \tildeScriptP {\ensuremath{\tilde{\cal P}}}
\def \twiddleScriptP {\tildeScriptP}  

\def \VistaApproximateDefiniteLuminosity {{927}}
\def \VistaApproximateLuminosity {{\ensuremath{\VistaApproximateDefiniteLuminosity}}}
\newcommand {\tabledSensitivityDiscussion}[1]{{}}

\def \numberOfVistaDistributions {{16,486}}
\def \numberOfVistaDiscrepantDistributions {{\ensuremath{384}}}
\def \numberOfVistaFinalStates {{344}}

\newcounter{mylistc}

\begin{document}

\title{Model-Independent Global Search for New High-$p_T$ Physics at CDF}
\affiliation{Institute of Physics, Academia Sinica, Taipei, Taiwan 11529, Republic of China} 
\affiliation{Argonne National Laboratory, Argonne, Illinois 60439} 
\affiliation{Institut de Fisica d'Altes Energies, Universitat Autonoma de Barcelona, E-08193, Bellaterra (Barcelona), Spain} 
\affiliation{Baylor University, Waco, Texas  76798} 
\affiliation{Istituto Nazionale di Fisica Nucleare, University of Bologna, I-40127 Bologna, Italy} 
\affiliation{Brandeis University, Waltham, Massachusetts 02254} 
\affiliation{University of California, Davis, Davis, California  95616} 
\affiliation{University of California, Los Angeles, Los Angeles, California  90024} 
\affiliation{University of California, San Diego, La Jolla, California  92093} 
\affiliation{University of California, Santa Barbara, Santa Barbara, California 93106} 
\affiliation{Instituto de Fisica de Cantabria, CSIC-University of Cantabria, 39005 Santander, Spain} 
\affiliation{Carnegie Mellon University, Pittsburgh, PA  15213} 
\affiliation{Enrico Fermi Institute, University of Chicago, Chicago, Illinois 60637} 
\affiliation{Comenius University, 842 48 Bratislava, Slovakia; Institute of Experimental Physics, 040 01 Kosice, Slovakia} 
\affiliation{Joint Institute for Nuclear Research, RU-141980 Dubna, Russia} 
\affiliation{Duke University, Durham, North Carolina  27708} 
\affiliation{Fermi National Accelerator Laboratory, Batavia, Illinois 60510} 
\affiliation{University of Florida, Gainesville, Florida  32611} 
\affiliation{Laboratori Nazionali di Frascati, Istituto Nazionale di Fisica Nucleare, I-00044 Frascati, Italy} 
\affiliation{University of Geneva, CH-1211 Geneva 4, Switzerland} 
\affiliation{Glasgow University, Glasgow G12 8QQ, United Kingdom} 
\affiliation{Harvard University, Cambridge, Massachusetts 02138} 
\affiliation{Division of High Energy Physics, Department of Physics, University of Helsinki and Helsinki Institute of Physics, FIN-00014, Helsinki, Finland} 
\affiliation{University of Illinois, Urbana, Illinois 61801} 
\affiliation{The Johns Hopkins University, Baltimore, Maryland 21218} 
\affiliation{Institut f\"{u}r Experimentelle Kernphysik, Universit\"{a}t Karlsruhe, 76128 Karlsruhe, Germany} 
\affiliation{Center for High Energy Physics: Kyungpook National University, Taegu 702-701, Korea; Seoul National University, Seoul 151-742, Korea; SungKyunKwan University, Suwon 440-746, Korea; Korea Institute of Science and Technology Information, Daejeon, 305-806, Korea; Chonnam National University, Gwangju, 500-757, Korea} 
\affiliation{Ernest Orlando Lawrence Berkeley National Laboratory, Berkeley, California 94720} 
\affiliation{University of Liverpool, Liverpool L69 7ZE, United Kingdom} 
\affiliation{University College London, London WC1E 6BT, United Kingdom} 
\affiliation{Centro de Investigaciones Energeticas Medioambientales y Tecnologicas, E-28040 Madrid, Spain} 
\affiliation{Massachusetts Institute of Technology, Cambridge, Massachusetts  02139} 
\affiliation{Institute of Particle Physics: McGill University, Montr\'{e}al, Canada H3A~2T8; and University of Toronto, Toronto, Canada M5S~1A7} 
\affiliation{University of Michigan, Ann Arbor, Michigan 48109} 
\affiliation{Michigan State University, East Lansing, Michigan  48824} 
\affiliation{University of New Mexico, Albuquerque, New Mexico 87131} 
\affiliation{Northwestern University, Evanston, Illinois  60208} 
\affiliation{The Ohio State University, Columbus, Ohio  43210} 
\affiliation{Okayama University, Okayama 700-8530, Japan} 
\affiliation{Osaka City University, Osaka 588, Japan} 
\affiliation{University of Oxford, Oxford OX1 3RH, United Kingdom} 
\affiliation{University of Padova, Istituto Nazionale di Fisica Nucleare, Sezione di Padova-Trento, I-35131 Padova, Italy} 
\affiliation{LPNHE, Universite Pierre et Marie Curie/IN2P3-CNRS, UMR7585, Paris, F-75252 France} 
\affiliation{University of Pennsylvania, Philadelphia, Pennsylvania 19104} 
\affiliation{Istituto Nazionale di Fisica Nucleare Pisa, Universities of Pisa, Siena and Scuola Normale Superiore, I-56127 Pisa, Italy} 
\affiliation{University of Pittsburgh, Pittsburgh, Pennsylvania 15260} 
\affiliation{Purdue University, West Lafayette, Indiana 47907} 
\affiliation{University of Rochester, Rochester, New York 14627} 
\affiliation{The Rockefeller University, New York, New York 10021} 
\affiliation{Istituto Nazionale di Fisica Nucleare, Sezione di Roma 1, University of Rome ``La Sapienza,'' I-00185 Roma, Italy} 
\affiliation{Rutgers University, Piscataway, New Jersey 08855} 
\affiliation{Texas A\&M University, College Station, Texas 77843} 
\affiliation{Istituto Nazionale di Fisica Nucleare, University of Trieste/\ Udine, Italy} 
\affiliation{University of Tsukuba, Tsukuba, Ibaraki 305, Japan} 
\affiliation{Tufts University, Medford, Massachusetts 02155} 
\affiliation{Waseda University, Tokyo 169, Japan} 
\affiliation{Wayne State University, Detroit, Michigan  48201} 
\affiliation{University of Wisconsin, Madison, Wisconsin 53706} 
\affiliation{Yale University, New Haven, Connecticut 06520} 
\author{T.~Aaltonen}
\affiliation{Division of High Energy Physics, Department of Physics, University of Helsinki and Helsinki Institute of Physics, FIN-00014, Helsinki, Finland}
\author{A.~Abulencia}
\affiliation{University of Illinois, Urbana, Illinois 61801}
\author{J.~Adelman}
\affiliation{Enrico Fermi Institute, University of Chicago, Chicago, Illinois 60637}
\author{T.~Akimoto}
\affiliation{University of Tsukuba, Tsukuba, Ibaraki 305, Japan}
\author{M.G.~Albrow}
\affiliation{Fermi National Accelerator Laboratory, Batavia, Illinois 60510}
\author{B.~\'{A}lvarez~Gonz\'{a}lez}
\affiliation{Instituto de Fisica de Cantabria, CSIC-University of Cantabria, 39005 Santander, Spain}
\author{S.~Amerio}
\affiliation{University of Padova, Istituto Nazionale di Fisica Nucleare, Sezione di Padova-Trento, I-35131 Padova, Italy}
\author{D.~Amidei}
\affiliation{University of Michigan, Ann Arbor, Michigan 48109}
\author{A.~Anastassov}
\affiliation{Rutgers University, Piscataway, New Jersey 08855}
\author{A.~Annovi}
\affiliation{Laboratori Nazionali di Frascati, Istituto Nazionale di Fisica Nucleare, I-00044 Frascati, Italy}
% ...
\collaboration{CDF Collaboration\footnote{With visitors from $^a$University of Athens, 15784 Athens, Greece, 
$^b$Chinese Academy of Sciences, Beijing 100864, China, 
$^c$University of Bristol, Bristol BS8 1TL, United Kingdom, 
$^d$University Libre de Bruxelles, B-1050 Brussels, Belgium, 
$^e$University of California, Irvine, Irvine, CA  92697, 
$^f$University of California Santa Cruz, Santa Cruz, CA  95064, 
$^g$Cornell University, Ithaca, NY  14853, 
$^h$University of Cyprus, Nicosia CY-1678, Cyprus, 
$^i$University College Dublin, Dublin 4, Ireland, 
$^j$University of Edinburgh, Edinburgh EH9 3JZ, United Kingdom, 
$^k$University of Heidelberg, D-69120 Heidelberg, Germany, 
$^l$Universidad Iberoamericana, Mexico D.F., Mexico, 
$^m$University of Manchester, Manchester M13 9PL, England, 
$^n$Nagasaki Institute of Applied Science, Nagasaki, Japan, 
$^o$University de Oviedo, E-33007 Oviedo, Spain, 
$^p$Queen Mary's College, University of London, London, E1 4NS, England, 
$^q$Texas Tech University, Lubbock, TX  79409, 
$^r$IFIC(CSIC-Universitat de Valencia), 46071 Valencia, Spain.
}}
\noaffiliation



\begin{abstract}
Data collected in Run II of the Fermilab Tevatron are searched for indications of new electroweak scale physics.  Rather than focusing on particular new physics scenarios, CDF data are analyzed for discrepancies with respect to the standard model prediction.  A model-independent approach (\Vista) considers the gross features of the data, and is sensitive to new large cross section physics.  A quasi-model-independent approach (\Sleuth) searches for a significant excess of events with large summed transverse momentum, and is particularly sensitive to new electroweak scale physics that appears predominantly in one final state.  This global search for new physics in over three hundred exclusive final states in \VistaApproximateLuminosity~pb$^{-1}$ of $p\bar{p}$ collisions at $\sqrt{s}=1.96$~TeV reveals no such significant indication of physics beyond the standard model. 
\end{abstract}

\maketitle
\clearpage


The particle physics standard model (SM) is remarkably successful, but is believed to require expansion beyond the electroweak scale.  A variety of possible extensions have been proposed.  Many analyses optimized for specific signatures have been performed to search for evidence of these possibilities.  Limits have been set on cross sections for postulated processes and on masses of hypothetical particles, but no conclusive indication of physics beyond the standard model has yet been seen~\cite{ReviewOfParticleProperties2006:Yao:2006px}.

This Letter summarizes a broad search for new physics at the electroweak scale without focusing on any specific proposed scenario.  The detailed writeup is provided in Ref.~\cite{VistaCdfPrd1}.  Events containing one or more particles produced at large transverse momentum collected by the CDF experiment in Run II of the Fermilab Tevatron are analyzed for discrepancies relative to the standard model prediction.  A model-independent approach (\Vista) considers gross features of the data, and is sensitive to new large cross section physics.  A quasi-model-independent approach (\Sleuth) emphasizes events with large summed scalar transverse momentum, and is particularly sensitive to new electroweak scale physics.  These global algorithms provide a complementary approach to searches optimized for more specific new physics scenarios.  Searches in a similar spirit have previously been performed by the \DZero\ Collaboration~\cite{SleuthPRL:Abbott:2001ke,SleuthPRD1:Abbott:2000fb,SleuthPRD2:Abbott:2000gx} in Tevatron Run I and by the H1 Collaboration~\cite{H1GeneralSearch:Aktas:2004pz} at HERA-I.


This search for new physics is designed with the intention of maximizing the chance for discovery, rather than excluding model parameter space if no discrepancy is found.  Discrepancies between data and a complete standard model background estimate are identified in a global sample of high transverse momentum (high-$p_T$) collision events.  Three statistics are employed to identify and quantify disagreement: populations of exclusive final states defined by the objects the events contain, shapes of kinematic distributions, and excesses on the tail of summed scalar transverse momentum distributions.  These statistics identify discrepancies worthy of further study.

A discovery claim can be made to the extent that a highlighted discrepancy can be demonstrated to be not due to a statistical fluctuation, a mismodeling of the detector response, or an inadequate implementation of the standard model prediction, and must therefore be due to some new underlying physics.  Any observed discrepancy is subject to scrutiny, and explanations are sought in terms of the above points.

The \Vista\ and \Sleuth\ algorithms provide a means for making the above three arguments, with a high threshold placed on the statistical significance of a discrepancy in order to minimize the chance of a false discovery  claim. As described later, this threshold is the requirement that the false discovery rate is less than 0.001, after taking into account the total number of final states, distributions, or regions being examined.

The traditional notions of signal and control regions are modified.  Removing prejudice as to where new physics may appear, all regions of the data are treated as both signal and control.  This analysis is not blind, but rather seeks to identify and understand discrepancies between data and the standard model prediction.  With the goal of discovery, emphasis is placed on examining discrepancies, focusing on outliers rather than global goodness of fit.  Individual discrepancies that are not statistically significant are generally not pursued.

\Vista\ and \Sleuth\ are employed simultaneously, rather than sequentially.  An effect highlighted by \Sleuth\ prompts additional investigation of the discrepancy, usually resulting in a specific hypothesis explaining the discrepancy in terms of a detector effect or adjustment to the standard model prediction that is then fed back and tested for global consistency using \Vista.

Forming hypotheses for the cause of specific discrepancies, implementing those hypotheses to assess their wider consequences, and testing global agreement after the implementation are emphasized as the crucial activities for the investigator throughout the process of data analysis~\footnote{It is not possible to systematically simulate the process of constructing, implementing, and testing hypotheses motivated by particular discrepancies, since this process is carried out by individuals.  The statistical interpretation of this analysis is made bearing this process in mind.}.  This process is constrained by the requirement that all adjustments be physically motivated.

This search for new physics terminates when one of two conditions are satisfied: either a compelling case for new physics is made, or there remain no statistically significant discrepancies on which a new physics case can be made. In the former case, to quantitatively assess the significance of the potential discovery, a full treatment of systematic uncertainties must be implemented. In the latter case, it is sufficient to demonstrate that all observed effects are not in significant disagreement with an appropriate global standard model description.

This analysis uses data corresponding to an integrated luminosity of \VistaApproximateLuminosity~pb$^{-1}$ of $p\bar{p}$ collisions at $\sqrt{s}=1.96$~TeV recorded by the CDF II detector~\cite{WandZCrossSectionPaper:Abulencia:2005ix}.  CDF II consists of a charged particle tracking system composed of silicon strip detectors and a gas drift chamber inside a 1.4~T magnetic field, surrounded by electromagnetic and hadronic calorimeters and enclosed by muon detectors.  

\begin{table}
\caption{A subset of the \Vista\ comparison between Tevatron Run II data and the standard model prediction, showing the final states with greatest discrepancies in population.  Final states are labeled in this table according to the number and types of objects present, and are ordered according to decreasing discrepancy between the total number of events expected and the total number observed in the data.  Only statistical uncertainties on the standard model prediction are shown; systematics are incorporated by allowing their values to float in the overall fit.  A total of 344 populated exclusive final states are considered.  
}
{\mbox{\hspace{0cm}\begin{minipage}{3.5in}
\tiny
\begin{tabular}{l@{ \ \ }r@{ \ \ \ }r@{ $\pm$ }lcl@{ \ \ }r@{ \ \ \ }r@{ $\pm$ }l}
\hline \hline 
{\bf Final State} & {\bf Data} & \multicolumn{2}{c}{{\bf SM prediction}} & & {\bf Final State} & {\bf Data} & \multicolumn{2}{c}{{\bf SM prediction}} \\ \hline 
3j$\tau^+$ & 71 & 114 & 4    & & e$^+\gamma$ & 636 & 551 & 11  \\
5j & 1661 & 1903 & 51    & &    e$^+$3j & 28656 & 27282 & 405 \\
2j$\tau^+$ & 233 & 297 & 6    & &    b5j & 131 & 95 & 5 \\
2j$2\tau^+$ & $6$ & $27$ & $4.6$ &  &    j2$\tau^+$ & 50 & 86 & 8 \\
be$^+$j & 2207 & 2015 & 29    & &    j$\tau^+\tau^-$ & 74 & 125 & 14 \\ 
3j  & 35436 & \ \ 37295 & 524    & &    b$p\!\!\!/$ & 10 & 30 & 5 \\ 
e$^+$3j$p\!\!\!/$ & 1954 & 1752 & 42    & &    e$^+$j$\gamma$ & 286 & 369 & 21 \\ 
be$^+$2j & 798 & 695 & 13    & &    e$^+$j$p\!\!\!/$$\tau^-$ & 29 & 14 & 2 \\ 
3j$p\!\!\!/$ & 811 & 968 & 38    & &    2j & 96502 & \ \ 92437 & 1355 \\ 
e$^+\mu^+$ & 26 & 12 & 2    & &     be$^+$3j & 356 & 299 & 8 \\ \hline \hline
\end{tabular}
\end{minipage}
}}
\label{tbl:vista_cdf}
\end{table}

\begin{figure}
\vspace*{-0.4cm}
\begin{tabular}{cc}
\includegraphics[width=3.5in]{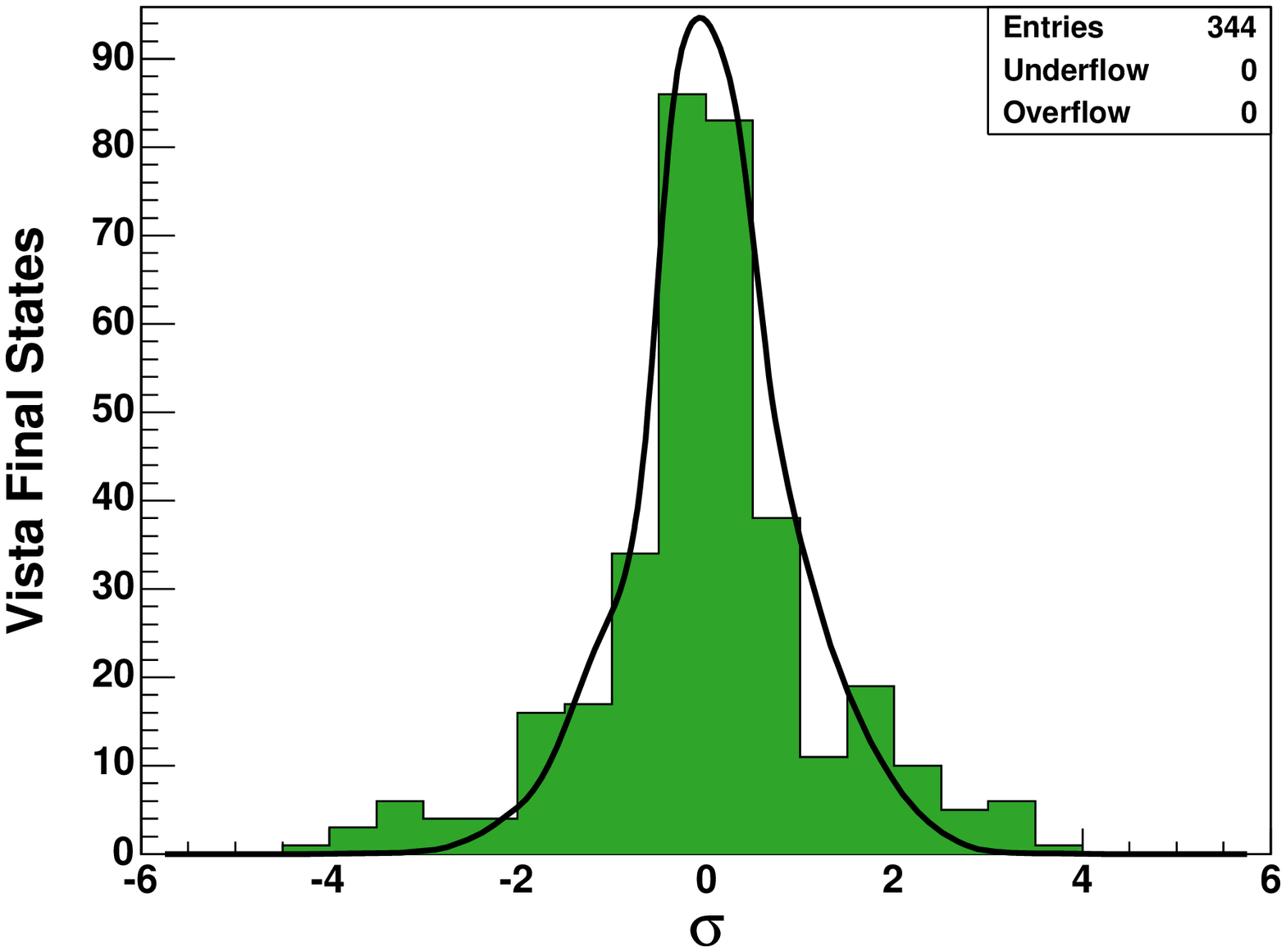} \\
\vspace*{-0.3cm}
\includegraphics[width=3.5in]{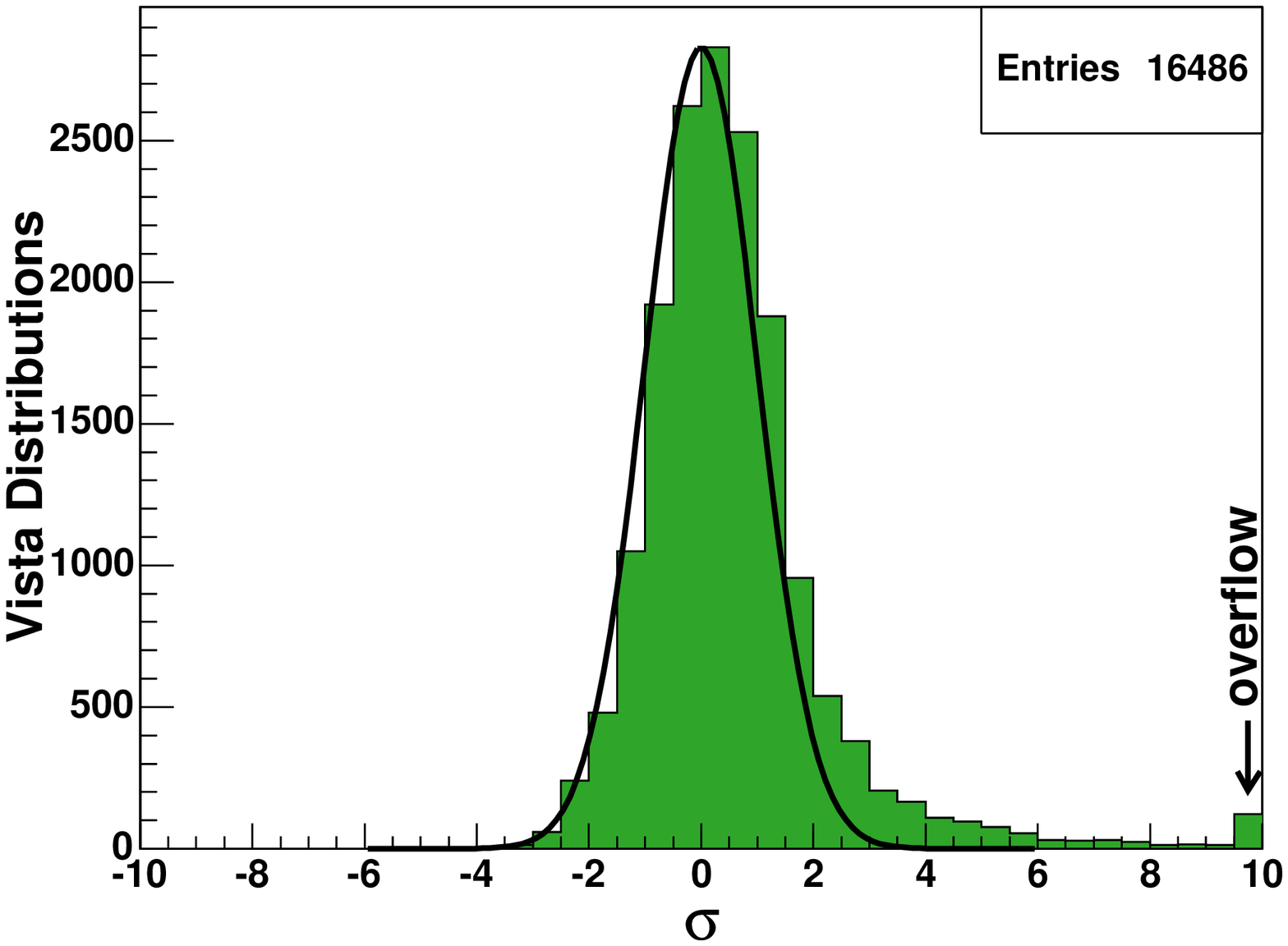} \\
\vspace*{-0.3cm}
\end{tabular}
\caption{Distribution of \Vista\ discrepancy between data and the standard model prediction, measured in units of standard deviation ($\sigma$), shown as the solid (green) histogram.  The top pane shows the distribution of discrepancies between the total number of events observed and predicted in the 344 populated final states considered.  The bottom pane shows the distribution of discrepancies between the observed and predicted shapes of 16,486 kinematic distributions.  In the bottom pane, distributions in which data and the standard model prediction are in agreement (large KS probability) correspond to negative $\sigma$, and distributions in which the data and the standard model prediction are in relative disagreement (small KS probability) correspond to large positive $\sigma$.  The expected distributions are shown as the solid (black) curves.  Interest is focused on the entries in the tails of the top distribution and the high tail of the bottom distribution.\vspace*{-0.3cm}}
\label{fig:VistaSummaryCdf}
\end{figure}

A standard set of object identification criteria is used to identify isolated and energetic objects produced in the hard collision, including electrons ($e^\pm$), muons ($\mu^\pm$), taus ($\tau^\pm$), photons ($\gamma$), jets ($j$), jets originating from a bottom quark ($b$), and missing momentum ($\pmiss$).  Monte Carlo event generators are used to determine the standard model prediction.  \Vista\ partitions data and Monte Carlo events into exclusive final states labeled according to the objects ($e^\pm$, $\mu^\pm$, $\tau^\pm$, $\gamma$, $j$, $b$, $\pmiss$) identified in each event.  Each event belongs to one and only one exclusive final state~\footnote{Events that are equivalent under global charge conjugation (such as a $e^+\pmiss$ event and a $e^-\pmiss$ event) are placed into a single final state (labeled in this case by $e^+\pmiss$).}.

A correction model is developed to improve systematic deficiencies in the standard model theoretical prediction and the simulation of the detector response.  Achieving this on the entire high-$p_T$ dataset requires a framework for quickly implementing and testing modifications to the correction model, including a quick fit for values of associated correction factors.  The specific details of the correction model are intentionally kept as simple as possible in the interest of transparency in the event of a possible new physics claim.  The details of this correction model are motivated by individual discrepancies noted in a global comparison of CDF high-$p_T$ data to the standard model prediction.  The correction model includes specific correction factors for the integrated luminosity of the sample, the ratio ($k$-factor) of the actual cross section for a standard model process and the usually leading order approximation given by event generators, object identification efficiencies, object misidentification rates, and trigger efficiencies.  A total of 44 correction factors are used, of which over twenty are constrained by external information.  A global $\chi^2$ is formed by comparison of CDF data to the standard model prediction, and minimized as a function of these correction factors.  Corrections to object identification efficiencies are typically less than 10\%; fake rates are consistent with an understanding of the underlying physical mechanisms responsible; $k$-factors range from slightly less than unity to greater than two for some processes with multiple jets.  

A global comparison of data to standard model prediction is made in \numberOfVistaDistributions\ kinematic distributions in \numberOfVistaFinalStates\ populated exclusive final states.  In each final state, the number of events observed is compared with the standard model prediction, as shown in Table~\ref{tbl:vista_cdf}, and the Poisson probability that the number of predicted events would fluctuate up to or above (or down to or below) the observed number of events is calculated and converted into units of standard deviation.  In each kinematic distribution, the shape of the data is compared to the shape of the standard model prediction using the Kolmogorov-Smirnov (KS) statistic, which is converted to a probability and then into units of standard deviation.  

\Vista\ highlights final states and kinematic distributions where the statistical significance of any discrepancy corresponds to a probability $< 0.001$ after accounting for the appropriate number of final states or distributions considered~\footnote{A probability of 0.001 after proper incorporation of the trials factor corresponds roughly to the usual criterion of $5\sigma$ if the trials factor is not accounted for.}.  The algorithm itself cannot determine whether a particular discrepancy constitutes a discovery of new physics.  Physics judgement is required to determine whether the discrepancy can be explained as a deficiency in the modeling of the CDF II detector response or in the calculation of the standard model prediction.

A summary of the \Vista\ comparison is shown in Fig.~\ref{fig:VistaSummaryCdf}.  The numbers of events observed are in agreement with the standard model prediction.  The narrow core of the histogram of \Vista\ final states (top of Fig.~\ref{fig:VistaSummaryCdf}) is due to final states with few data events.  The excess at large $\sigma$ in the histogram of \Vista\ distributions (bottom of Fig.~\ref{fig:VistaSummaryCdf}) shows disagreement between data and standard model prediction in some distributions.  The number of distributions showing a significant ($>3\sigma$ after the trials factor) difference in shape between data and the standard model prediction is \numberOfVistaDiscrepantDistributions.  Of these, 312 are attributed to modeling the parton radiation (with 186 of these 312 pointing out that individual jet masses are larger in data than in the prediction), and 59 reflect an inadequate modeling of the overall transverse boost of the system (``intrinsic $k_T$'').  The nature of these discrepant distributions makes it difficult to use them to support a new physics claim, since at present these discrepancies appear most probably due to an imperfect implementation of the standard model prediction.  Further investigation into obtaining an adequate QCD-based description is continuing.  The remaining 13 discrepant distributions arise from the coarseness of the correction model.  Additional details are provided in Ref.~\cite{VistaCdfPrd1}.


\begin{figure}
\includegraphics[width=2.6in,angle=270]{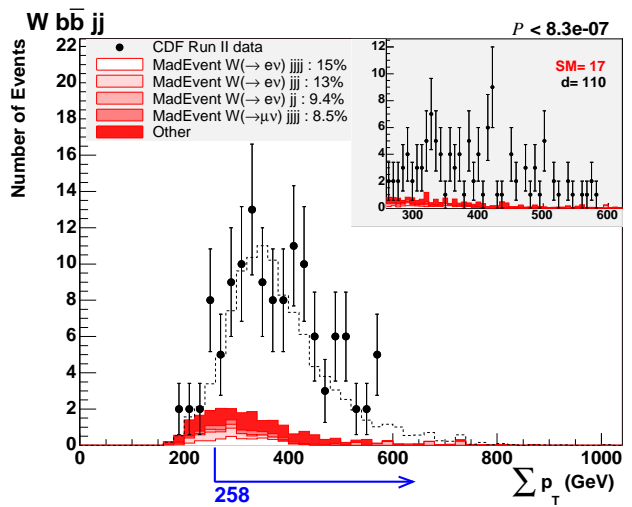}
\caption{A test to see whether \Sleuth\ would find evidence of top quark pair production, if the top quark were not known.  Shown is the \Sleuth\ $W b\bar{b} jj$ final state, consisting of events with one electron or muon, missing transverse energy, and $\geq4$~jets, at least one of which is $b$-tagged. The CDF Run II data are shown as (black) filled circles, the standard model prediction (minus $t\bar{t}$ production) is shown as the (red) shaded stacked histogram, and the expected contribution from $t\bar{t}$ is shown as the dashed line.  \Sleuth\ chooses the region with $\SumPt>258$~GeV, shown by the (blue) arrow, and displayed in the inset.  In this region, the number of predicted standard model events ({\em{sans}} $t\bar{t}$) is $\text{SM}=17$, and the number of observed data events is $d=110$. \Sleuth\ quantifies the fraction of hypothetical similar experiments that would have produced a region more interesting than the region chosen in this final state, and finds $\scriptP_{Wb\bar{b}jj}<8.3\times10^{-7}$, corresponding to a value of $\twiddleScriptP$ that easily satisfies \Sleuth's discovery threshold of $\twiddleScriptP < 0.001$.}
\label{fig:topless_SM_sensitivityTest}
\end{figure}

\Sleuth~\cite{SleuthPRL:Abbott:2001ke,SleuthPRD1:Abbott:2000fb,SleuthPRD2:Abbott:2000gx,H1GeneralSearch:Aktas:2004pz,ACAT2003:Knuteson:2004nj} is simultaneously used to search for evidence of new physics on the high-$p_T$ tails.  \Sleuth\ is a quasi-model-independent search technique, based on the assumption that new electroweak-scale physics will manifest itself as a high-$p_T$ excess of data over the standard model expectation in a particular final state.  The strengths and limitations of \Sleuth\ follow directly from these assumptions.

\Sleuth\ considers a single variable, the summed scalar transverse momentum (\SumPt) of all objects in the event. The standard model prediction for the distribution of \SumPt\ is determined using the correction factors found by \Vista.  For each final state, \Sleuth\ determines the most interesting region on the tail of this distribution.  A final state contains as many regions as data points, where the $d^{\text{th}}$ region is defined as the semi-infinite interval with lower bound equal to the $d^{\text{th}}$ largest data \SumPt.  The $d^{\text{th}}$ region contains $d$ data events; the number of events expected from the standard model is obtained by integrating the predicted standard model \SumPt\ distribution over this semi-infinite region.
 
For a region containing $d$ data points, $p_d$ is defined as the Poisson probability that the standard model prediction would fluctuate up to or above $d$.  The most interesting region $\mathcal{R}$ is defined as the region for which $p_d$ is smallest.  Pseudo experiments are performed by drawing pseudo data from the standard model \SumPt\ distribution, and the most interesting region is found for each pseudo experiment. The fraction \scriptP\ of these pseudo experiments producing a region more interesting than the region $\mathcal{R}$ found in the data quantifies the interest of this final state.  


Considering all final states, \Sleuth\ determines the most interesting final state in the CDF high-$p_T$ data, and calculates \tildeScriptP, the fraction of hypothetical similar CDF experiments that would have produced a region in any final state more interesting than the most interesting region in the most interesting final state.  In calculating \tildeScriptP, \Sleuth\ rigorously accounts for the number of final states that have been considered.  With an accurate correction model and in the absence of new physics, the distribution of \tildeScriptP\ is uniform between zero and unity; in the presence of new physics, small \tildeScriptP\ is expected.  The threshold for pursuit of a possible discovery case is taken to be $\tildeScriptP < 0.001$.

Figure~\ref{fig:topless_SM_sensitivityTest} shows a sensitivity test in which the standard model process $p\bar{p}\rightarrow t\bar{t}$ is subtracted from the standard model background and observed as an excess in the CDF data.  \Sleuth\ observes the top quark with an integrated Run II luminosity comparable to that accumulated by CDF and \DZero\ in Tevatron Run I when the top quark discovery was announced~\cite{CDFTopDiscovery:Abe:1995hr1,D0TopDiscovery:Abachi:1995iq1}.  Several other sensitivity tests have been conducted with pseudo signal events injected into pseudo data drawn from the standard model prediction.  On these sensitivity tests, \Sleuth\ performs comparably to targeted searches for phenomena satisfying \Sleuth's basic assumptions that new physics will appear as an excess of data over the standard model prediction at large summed scalar transverse momentum in one primary final state.

In \VistaApproximateDefiniteLuminosity~pb$^{-1}$ of CDF Run II data, \Sleuth\ finds $\twiddleScriptP = 0.46$.  Assuming any deficiencies in the standard model implementation and detector simulation are accurately resolved by the correction model, the fraction of hypothetical similar CDF experiments that would observe something as interesting as the most interesting region observed in the CDF Run II data is 46\%.  None of the regions examined surpass \Sleuth's discovery threshold.  Further discussion of the most interesting regions is provided in Ref.~\cite{VistaCdfPrd1}.


In conclusion, a broad search for new physics (\Vista) has been performed in \VistaApproximateLuminosity~pb$^{-1}$ of CDF Run II data.  A complete standard model background estimate has been obtained and compared with data in 344 populated exclusive final states and 16,486 relevant kinematic distributions.  Consideration of exclusive final state populations yields no statistically significant ($>3\sigma$) discrepancy after the trials factor is accounted for.  Quantifying the difference in shape of kinematic distributions using the Kolmogorov-Smirnov statistic, significant discrepancies are observed between data and standard model prediction.  These discrepancies are believed to arise from mismodeling of the parton shower and intrinsic $k_T$, and represent observables for which a QCD-based understanding is highly motivated.  None of the shape discrepancies highlighted motivates a new physics claim.

A further systematic search (\Sleuth) for regions of excess on the high-$\SumPt$ tails of exclusive final states has been performed, representing a quasi-model-independent search for new electroweak scale physics.  A measure of interest rigorously accounting for the trials factor associated with looking in many regions is defined, and used to quantify the most interesting region observed in the CDF Run II data.  No region of excess on the high-$\SumPt$ tail of any of the \Sleuth\ exclusive final states surpasses the discovery threshold.

Although this global analysis of course cannot prove that no new physics is hiding in these data, this broad search of the Tevatron Run II data represents one of the single most encompassing tests of the particle physics standard model at the energy frontier.

\acknowledgments
We thank the Fermilab staff and the technical staffs of the participating 
institutions for their vital contributions. This work was supported by 
the U.S. Department of Energy and National Science Foundation; the 
Italian Istituto Nazionale di Fisica Nucleare; the Ministry of Education, 
Culture, Sports, Science and Technology of Japan; the Natural Sciences 
and Engineering Research Council of Canada; the National Science Council 
of the Republic of China; the Swiss National Science Foundation; the A.P. 
Sloan Foundation; the Bundesministerium f\"ur Bildung und Forschung, 
Germany; the Korean Science and Engineering Foundation and the Korean 
Research Foundation; the Particle Physics and Astronomy Research Council 
and the Royal Society, UK; the Institut National de Physique Nucleaire et 
Physique des Particules/CNRS; the Russian Foundation for Basic Research; 
the Comisi\'on Interministerial de Ciencia y Tecnolog\'{\i}a, Spain;
the European Community's Human Potential Programme; the Slovak R\&D Agency; and
the Academy of Finland.

\bibliography{prl1}

\end{document}